\documentclass[runningheads]{llncs}

\usepackage{graphicx}
\usepackage[dvipsnames]{xcolor}
\usepackage{amsmath}
\usepackage{amsfonts,epsfig,latexsym,amssymb,yhmath}
\usepackage[pdftex]{hyperref}

\usepackage{booktabs}
\usepackage{array}
\newif\ifdraft
\drafttrue
\ifdraft

\def\Hc#1{\textcolor{Aquamarine}{\textit{\textsf{ \small [H: #1]}}}}  % <----- COMMENT
\def\Hd#1{\textcolor{purple}{\textit{[deleted: #1]}}}  % <----- DELETED

\def\Ac#1{\textcolor{Orange}{\textit{\textsf{ \small [A: #1]}}}}  % <----- COMMENT
\def\Ad#1{\textcolor{Red}{\textit{[deleted: #1]}}}  % <----- DELETED

\def\Wc#1{\textcolor{ForestGreen}{\textit{\textsf{ \small [W: #1]}}}}  % <----- COMMENT
\def\Wd#1{\textcolor{purple}{\textit{[deleted: #1]}}}  % <----- DELETED

\def\Sc#1{\textcolor{Mulberry}{\textit{\textsf{ \small [S: #1]}}}}  % <----- COMMENT
\def\Sd#1{\textcolor{Mulberry}{\textit{[deleted: #1]}}}  % <----- DELETED

\else

\def\Hc#1{}  % <----- COMMENT
\def\Hd#1{}  % <----- DELETED

\def\Ac#1{}  % <----- COMMENT
\def\Ad#1{}  % <----- DELETED

\def\Wc#1{}  % <----- COMMENT
\def\Wd#1{}  % <----- DELETED

\def\Sc#1{}  % <----- COMMENT
\def\Sd#1{}  % <----- DELETED
\fi

\DeclareMathAlphabet{\mathbfsf}{\encodingdefault}{\sfdefault}{bx}{n}
\newcommand{\qvec}[1]{{\mathit{\boldsymbol{#1}}}}

\begin{document}
\title{ Modeling Disease Progression In Retinal OCTs With Longitudinal Self-Supervised Learning}

\author{Antoine Rivail \inst{1} \and Ursula Schmidt-Erfurth
\inst{1} \and Wolf-Dieter Vogl \inst{1} \and Sebastian M. Waldstein \inst{1} \and Sophie Riedl \inst{1} \and Christoph Grechenig \and \inst{1} Zhichao Wu \inst{2,3}  \and Hrvoje Bogunovi\'c \inst{1}}

\institute{ Christian Doppler Laboratory for Ophthalmic Image Analysis, Department of Ophthalmology and Optometry, Medical University of Vienna, Austria \and  Centre for Eye Research Australia, Royal Victorian Eye and Ear Hospital, East Melbourne, Australia \and Ophthalmology, Department of Surgery, The University of Melbourne, Melbourne, Australia}

\authorrunning{A. Rivail et al.}

\titlerunning{Disease progression in retinal OCT with self-supervised learning}
% If the paper title is too long for the running head, you can set
% an abbreviated paper title here
%

\maketitle              % typeset the header of the contribution
\begin{abstract}
Longitudinal imaging is capable of capturing the static ana\-to\-mi\-cal structures and the dynamic changes of the morphology resulting from
aging or disease progression. 
Self-supervised learning allows to learn new representation from available large unlabelled data without any expert knowledge. 
We propose a deep learning self-supervised approach to model disease progression from longitudinal retinal optical coherence tomography (OCT). Our self-supervised model takes benefit from a generic time-related task, by learning to estimate the time interval between pairs of scans acquired from the same patient. This task is (i) easy to implement, (ii) allows to use irregularly sampled data, (iii) is tolerant to poor registration, and (iv) does not rely on additional annotations. This novel method learns a representation that focuses on progression specific information only, which can be transferred to other types of longitudinal problems. We transfer the learnt representation to a clinically highly relevant task of predicting the onset of an advanced stage of age-related macular degeneration within a given time interval based on a single OCT scan. The boost in prediction accuracy, in comparison to a network learned from scratch or transferred from traditional tasks, demonstrates that our pretrained self-supervised representation learns a clinically meaningful information.

\end{abstract}
\section{Introduction}
Due to a rapid advancement of medical imaging, the amount of longitudinal imaging data is rapidly growing~\cite{Fujimoto_2016}. Longitudinal imaging is an especially effective observational approach, used to explore how disease processes develop over time in a number of patients, providing a good indication of disease progression. It enables personalized precision medicine~\cite{Vogl_2017} and it is a great source for automated image analysis. 
However, the automated modelling of disease progression faces many challenges: despite the large amount of data, associated human-level annotations are rarely available, which leads to several limitations in the current modelling methods. They are either limited to the annotated cross-sectional samples and miss most of the temporal information. Or, they are based on known handcrafted features and simplified models of low complexity. 

To overcome these limitations, we propose a solution based on self-supervised learning. Self-supervised learning consists of learning an auxiliary or a so-called \emph{pretext} task on a dataset without the need for human annotations to generate a generic representation. This representation can then be transferred to solve complex supervised tasks with a limited amount of data. We propose a pretext task that exploits the availability of large numbers of unlabelled longitudinal images, focuses on learning temporal-specific patterns and that is not limited by irregular time-sampling or lack of quality in image registration, which are common issues in longitudinal datasets. 
The learnt representation is compact and allows for transfer learning to more specific problems with limited amount of data. We  demonstrate  the  capability  of  our  proposed  method  on  a longitudinal  retinal  optical  coherence  tomography  (OCT) dataset of patients with early/intermediate age-related macular degeneration (AMD).
\paragraph{Clinical background} In the current ophthalmic clinical practice, optical coherence tomography (OCT) is the most commonly used retinal imaging modality. It provides 3-dimensional in-vivo information of the (pathological) retina with a micrometer resolution. Typically, volumetric OCTs are rasterized scans, where each \emph{B-Scan}  is a cross-sectional image of the retinal morphology.
AMD is a major epidemic among the elderly and advanced stage of AMD is the most common cause of blindness in industrialized countries, with two identified main forms: geographic atrophy~(GA) and choroidal neovascularization~(CNV). The progression from early or \emph{intermediate}, symptomless stages of AMD to \emph{advanced} stage is extremely variable between patients and very difficult to estimate clinically. Robust and accurate prediction at the individual patient level is a critically important medical need in order to manage or prevent irreversible vision loss. 

\paragraph{Related work} Current analysis of clinical data and disease course development evolves around traditional statistical approaches, where time-series models are fit to a limited amount of known biomarkers describing the disease status~\cite{Vogl_2017,Vogl2017a}. Such models are problematic in case of a disease such as AMD where the underlying mechanisms are still poorly understood and main biomarkers are yet undiscovered~\cite{schmidt-erfurth_prediction_2018}. Disease progression modelling from longitudinal imaging data has been most active in the field of Neuroimaging for modeling the progression of Alzheimer’s disease, largely due to the public availability of a longitudinal brain magnetic resonance images under Alzheimer’s Disease Neuroimaging Initiative (ADNI). There, a variety of regression-based methods have been applied to fit logistic or polynomial functions to the longitudinal dynamic of each imaging biomarker \cite{sabuncu_event_2014}. Other efforts have been focusing on non-parametric Gaussian process (GP) models~\cite{lorenzi_efficient_2015} but the specification of the joint covariance structure of the image features to account for spatial and temporal correlation has been found to still be computationally prohibitive. In addition, these methods are often linear, and/or treat the data as cross-sectional, and thus do not exploit non-linear relationships. Self-supervised learning was already successfully applied to time-series video data in the field of computer vision, where Lee et al. developed a solution based on time-shuffling \cite{DBLP:journals/corr/abs-1708-01246}, which inspired our method.

\paragraph{Contribution}
%Contribuion
We propose a novel self-supervised task, suited for learning a compact representation of longitudinal imaging data that captures time-specific patterns without the need of segmentation or a priori information. We assume that the information of the future evolution and disease progression is encoded in an observed series of images to a certain degree. Hence, we train our model on a pretext task: estimating the time interval between pairs of images of the same patient. Thus, an implicit aging model is built resulting in a compact representation that contains knowledge about healthy and disease evolution. As this task does not rely on annotations, perfect registration or regular sampling intervals, we are able to incorporate large unlabelled longitudinal datasets without the need of time and cost intensive pre-processing or annotation generation. We demonstrate that the model is able to learn the given pretext task and that it is capable of capturing the longitudinal evolution. Furthermore, we show that such a representation can be transferred to other longitudinal problems such as a prediction or survival estimation setting with limited amount of training data. In our case, we predict the future conversion to advanced stage of AMD within a certain time interval. In contrast to a model learnt from scratch or transferred from non longitudinal task (ex. autoencoder), we observe a boost in accuracy when fine-tuning a model trained on our new pretext task.  

%
%
%

%%%%%%%%%%%%%%%%%%%%%%%%%%%%%%%%%%%%%%%%%%%%%%%%%%%%%%%%%%%%%%%%%%%%%%%55
\section{Self-supervised learning of spatio-temporal representations}
%Self-supervised learning, a type of unsupervised learning, consists of learning a pretext task that can be derived directly from the acquired data without %any additional information. 
%Most of this tasks consist of reconstructing intentionally corrupted data. 
%A deep network trained on such tasks learns representation that can be rapidly transferred to new tasks. Thus the task should be generic enough to capture %a multi-purpose representation. By being self-supervised, this method allows to process a large amount of unlabelled data to learn such a representation. 

In this Section, we will present the pretext task that we chose for self-supervised learning of longitudinal imaging data, the deep networks that we implemented to solve it, and finally how we extract the representations to transfer them to different problems.

\begin{figure}[tb]
\centering
\includegraphics[width=\textwidth]{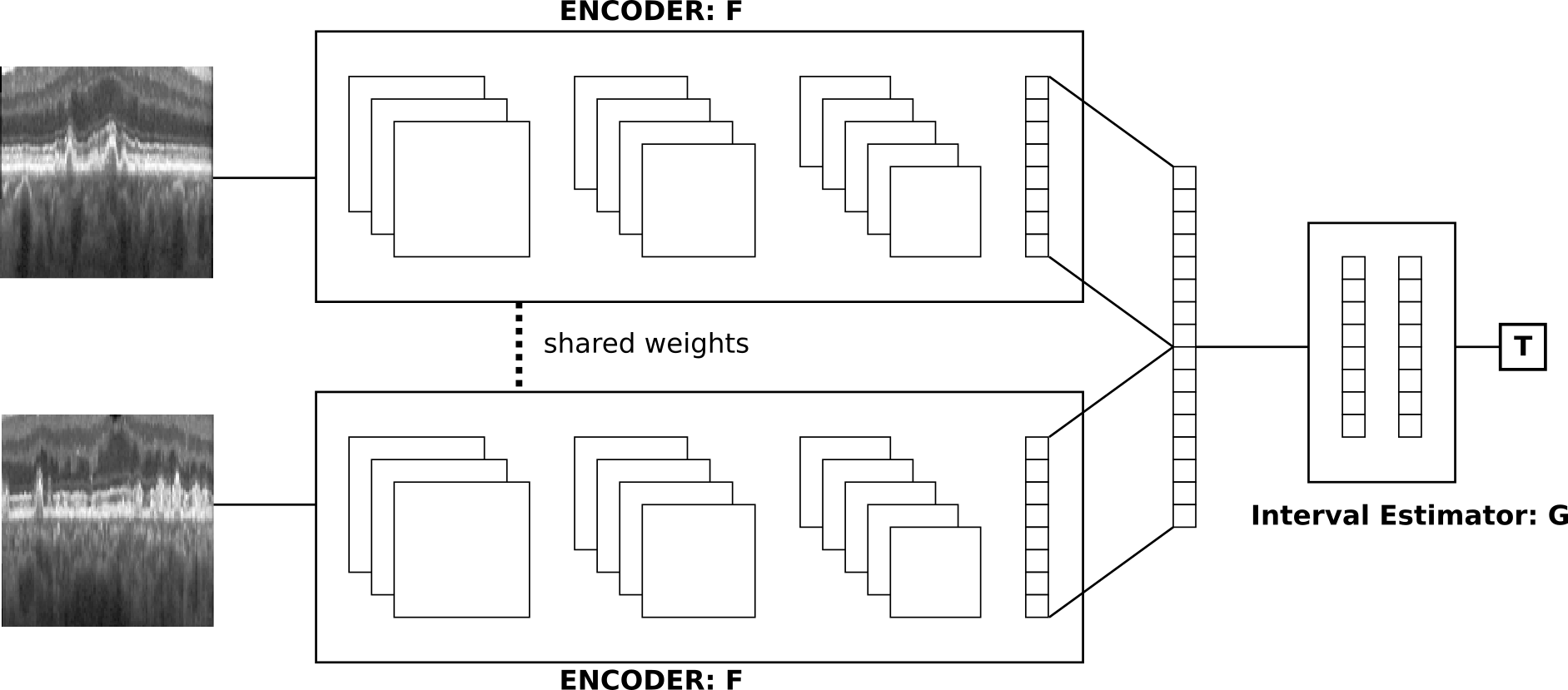}
\caption{Deep siamese network learning the self supervised task. It consists of predicting the acquisition time interval separating a pair of OCT scans of the same patient. Each B-Scan is encoded by the \textbf{encoder}, then the codes are used as input for the \textbf{pair-interval} network, which outputs the estimated time interval.}
\label{figm0}
\end{figure}

\paragraph{Self-supervised learning paradigm}
to learn the pretext task, we train a deep Siamese network (Fig.~\ref{figm0})~\cite{bromley1994signature}. The Siamese structure reflects the symmetry of the problem and allows to train a single encoder, which can be later transferred either as a fixed feature extractor or as a pretrained network for fine-tuning. 

Let $X_{t_1}$, $X_{t_2}$ be two OCT images acquired from the same patient, and $t_1$, $t_2$ be the corresponding time-points of acquisition. These images are encoded into a compact representation ($H_{t_1}$, $H_{t_2}$) by the \textit{encoder network}, $F$. The \textit{pair-interval network}, $G$, predicts the time interval or relative time difference between the pair of B-Scans, $\Delta T = t_2 - t_1 $. Note that the order of the pair do not need to be chronological. 

\begin{gather}
\operatorname{F:}\;  \qvec{X}_{t} \rightarrow \qvec{H_t} \\
\operatorname{G:}\;  [\qvec{H}_{t_1},\qvec{H}_{t_2}] \rightarrow \widehat{\qvec{\Delta T}} \\
%\qvec{\Delta T} = t_2-t_1 \\
\operatorname{Loss} = || \operatorname{G}( [\qvec{H}_{t_1},\qvec{H}_{t_2}]) - \qvec{\Delta T}  ||_2 
\end{gather}
The entire siamese network is trained by minimizing the $L_2$ loss of this regression task. After the training, the encoder network can be transferred to extract features for other tasks.%

\paragraph{Implementation} the encoder is implemented as a deep convolutional network, with three blocks of three layers (each layer: 3x3 convolution layer with batch normalization and ReLU activation with 16,32,64 channels for block 1,2,3) with a max pooling layer at the end of each block. The last block is followed by a fully connected layer (128 units), which outputs the encoded version of the B-scan (denoted as \textbf{vgg}). We also tested a version with skip connections followed by concatenation between the blocks (denoted as \textbf{dense}). The  pair-interval network has two fully connected layers and outputs the estimated time interval. 

\paragraph{Learning setup} The network is trained by minimizing L2 loss with gradient-descent algorithm Adam \cite{kingma2014adam}. We trained the network for 600k steps and computed validation loss every 12k steps. We kept the model with highest validation loss.

\begin{figure}[tb]
\centering
\includegraphics[width=\textwidth]{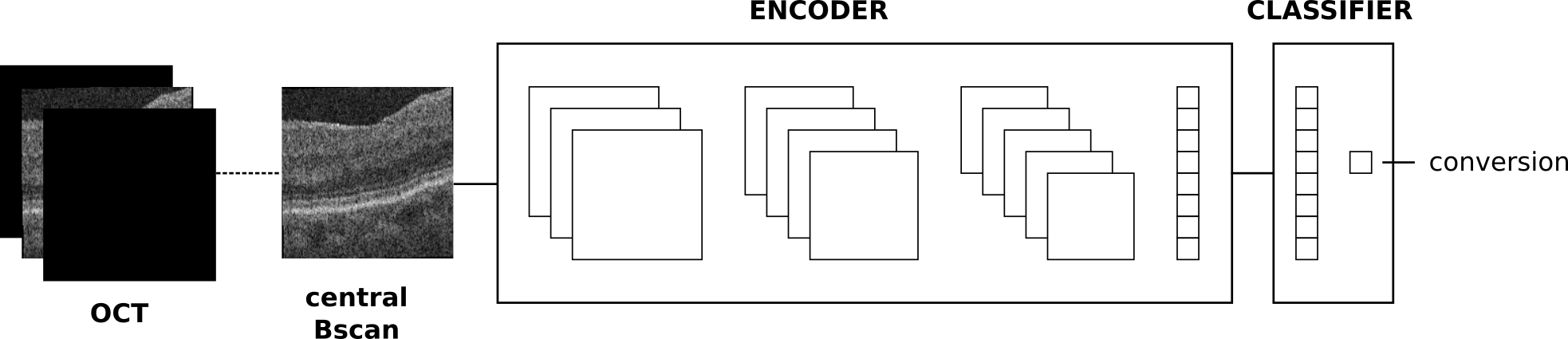}
\caption{Transfer learning of the trained encoder. A classification block is added at the end of the encoder and the network is fine-tuned on the new problem.} \label{figm2}
\end{figure}

\paragraph{Transfer of representations} The representations, $\mathbf{H}_t$, extracted from the encoder network, are used as input for a classification problem. This transfer allows to evaluate whether these representations are containing meaningful information regarding the patient-specific evolution of AMD. We directly transfer the trained encoder from the deep siamese network to a classification task by adding a final block to perform classification (Fig ~\ref{figm2}). The classification block consists of two fully connected layers, the first one with ReLU activation, the last one with softmax. The resulting network is fine-tuned by minimizing cross-entropy.

%%%%%%%%%%%%%%%%%%%%%%%%%%%%%%%%%%%%%%%%%%%%%%%%%%%%%%%%%%%%%%%%%%%%%%%55
\section{Experiments and Results}
Here, we provide details about the self-supervised training and its evaluation with respect to the time prediction. Then, we transfer the representation obtained from self-supervised training to a classification task, where we predict from longitudinal retinal OCT the conversion from intermediate to advanced AMD within different time intervals. 

\subsection{Dataset}
The longitudinal dataset used for training and validation contains 3308 OCT scans from 221 patients (420 eyes) diagnosed with intermediate AMD. Follow-up scans were acquired in a three or six months interval up to 7 years, and were included in the dataset up to the time-point of conversion to advanced AMD. Follow-up acquisitions were automatically registered by scanner software (Spectralis OCT \textsuperscript{\textregistered}, Heidelberg Engineering, GER). Within this population, 48 eyes converted to GA, an advanced stage of AMD. 
\paragraph{Cross-validation} The patients are divided in 6 fixed folds to perform cross-validation. For the pretext task and the prediction of conversion, we used one fold fold as test, one as validation and the remaining folds as training data. The pretraining with self-supervised learning use the same training sets as the prediction of conversion.

\paragraph{Preprocessing} First, the bottom-most layer of the retina, the Bruch's membrane (BM), was segmented using \cite{chen_three-dimensional_2012}, followed by a flattening of the concave structure and an alignment of the BM over all scans. Finally, scans were cropped to the same physical field of view (6 mm $\times$ 0.5 mm) and resampled to $128 \times 128$ .

\subsection{Learning the pretext task}

\begin{figure}[tb]
\centering
\includegraphics[width=\textwidth]{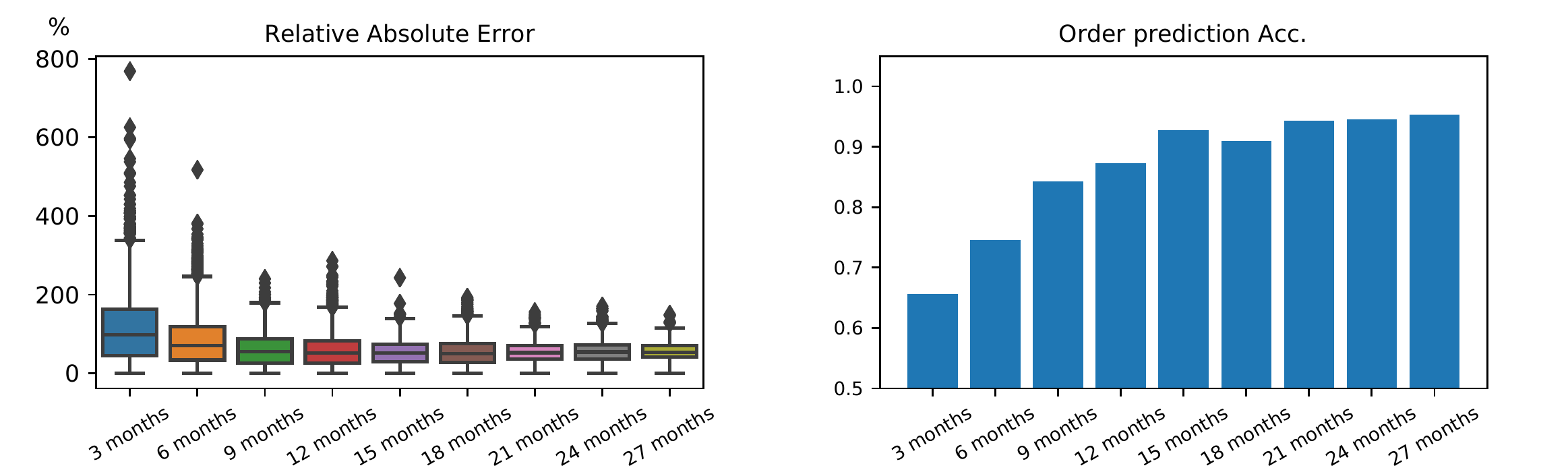}
\caption{\textit{Left}, boxplot of relative absolute error $[\%]$ for the different time intervals in the dataset (3 to 24 months).\textit{Right}, sample order prediction accuracy for the different time intervals in the dataset.} \label{boxplot}
\end{figure}

The success of learning the pretext task was evaluated using R$^2$ and the mean absolute error (MAE) of the \textit{interval} prediction. In addition, we verified how well the network could predict the temporal \textit{order} of samples, which is done by evaluating the accuracy of predicting the correct sign of the interval. To obtain a volume prediction (the network is trained on B-Scans), we took the mean of all scan predictions.  The best performance was achieved by the vgg-like model with a R$^2$ of \textbf{0.566}, a MAE of \textbf{7.69 months} and an accuracy for order prediction of \textbf{0.843} (See Table 1).  
Fig.~\ref{boxplot}  displays MAE and order prediction accuracy for the different time intervals (the visit intervals are roughly a multiple of three months). We observed that the network was able to predict the order, even for the smallest interval (3 months) with an accuracy of \textbf{0.66} (a random performance yielding 0.5). The absolute time-interval regression error was greater for large intervals (Fig ~\ref{boxplot}), with a tendency to underestimate the interval, probably because of a uniform distribution of training intervals centered on zero. However the relative error was decreasing for larger intervals.
These results show that it is possible to estimate the time interval between two OCTs to a certain extent, which allows to learn a generic evolution model for the retina. In the next experiment, we verified that this model contained relevant longitudinal information to solve a specific clinical prediction task.

\begin{table}[tb]
\centering
\caption{Evaluation of interval prediction (R$^2$ and MAE), and of sample order prediction (accuracy) in the pretext task obtained from a 6-fold cross-validation.}\label{tab1}
\begin{tabular}{llcl}\toprule
Network &  R$^2$ & MAE (months)& Accuracy \\
\midrule
vgg &  \textbf{0.566} & \textbf{7.69} & \textbf{0.843}   \\
dense &  0.505 & 8.35 &0.823  \\
\bottomrule
\end{tabular}
\end{table}

\subsection{Conversion to advanced AMD classification }
We applied the representation on a binary classification task, where we predicted from a single OCT representation, whether a patient eye will convert to GA within defined intervals of 6 months, 12 and 18 months (three separated binary problems). For the evaluation dataset we used the same 6 folds and their train, validation and test subsets. However, we included only one OCT per patient, in order to simulate a single visit. For patients who converted we chose the acquisition having the largest distance to conversion within the given interval. For patients showing no conversion within the study we chose the acquisition within the given interval with the furthest distance to the last patients acquisition. For each OCT volume, we used the trained encoder to extract a representation vector from the central B-Scan.

We restricted the input to a single time-point to verify that the encoded information allows to evaluate directly the stage of the patient. We fine-tuned the encoder on each fold and kept the epoch with the highest validation loss. We tested two different baselines with the same structure as the model transferred from the self-supervised learning, (i) trained from scratch (no transfer), and (ii) transferred from an autoencoder trained on cross-sectional OCTs with mean square error as reconstruction loss. The Autoencoder baseline allows to verify that our method is learning specific longitudinal features. The networks pretrained with our method or with an autoencoder are fine-tuned using ADAM optimizer by minimizing cross-entropy. After 20 epochs, the best epoch is selected using validation AuC. We performed a grid search on the learning rate, number of features in the classification block, and dropout rate, each setting is repeated 5 times. The best hyperparameter combination was chosen based on the best average validation AuC. The final cross-validated test performance was evaluated using ROC AuC and average precision. We observed that all the networks rapidly overfit on the dataset, which was expected given the size of the dataset (around 260 training samples) and the high capacity of the network. For all intervals, the self-supervised method shows best performance for both ROC AuC and average precision. On the other hand, transfer from the OCT autoencoder is only marginally better than the network trained from scratch (Table ~\ref{tabtransfer}). The difference between the transfer from OCT autoencoder and our new self-supervised task shows that the latter captures longitudinal information, which is not available in the autoencoder. Our method allows to quickly train a deep network on this challenging task with a small number of annotations.

\begin{table}[tb]
\setlength{\tabcolsep}{6pt}
\centering
\caption{Conversion classification of patients suffering from intermediate AMD. We performed a 6-fold cross-validation and display the ROC AuC and average precision (mean and standard deviation) for three settings: conversion within 6 months (m.), 12 months and 18 months.}
\begin{tabular}{llll}\toprule
\setlength{\tabcolsep}{12pt}
\def\arraystretch{1.5}
&  \multicolumn{3}{c}{ROC AuC} \\
Model &  6 m. & 12 m. & 18 m.  \\
\midrule
Training from scratch (i)               & 0.640 $\pm$ 0.067         & 0.651 $\pm$ 0.076          & 0.676 $\pm$
0.095  \\
OCT autoencoder (ii)                        & 0.650 $\pm$ 0.144          & 0.519 $\pm$ 0.060            &0.677 $\pm$ 0.088           \\
Self-supervised (ours)   & \textbf{0.753}$\pm$ 0.061     & \textbf{0.784} $\pm$ 0.067 &  \textbf{0.773} $\pm$ 0.074   \\
\midrule
&  \multicolumn{3}{c}{Average Precision} \\
Model &  6 m. & 12 m. & 18 m.  \\
\midrule
Training from scratch (i)               &            0.309 $\pm$ 0.114       & 0.282 $\pm$ 0.0125         & 0.300 $\pm$ 0.107 \\
OCT autoencoder (ii)                      &0.277 $\pm$ 0.152                  &0.283 $\pm$ 0.144              &0.329 $\pm$ 0.117   \\
Self-supervised (ours)    &  \textbf{0.367 }$\pm$ 0.084  & \textbf{0.394} $\pm$  0.115 & \textbf{0.463} $\pm$ 0.133 \\

\bottomrule
\end{tabular}
\label{tabtransfer}
\end{table}

%%%%%%%%%%%%%%%%%%%%%%%%%%%%%%%%%%%%%%%%%%%%%%%%%%%%%%%%%%%%%%%%%%%%%%%55
\section{Discussion and conclusions}
Effective modeling of disease progression from longitudinal data has been a long pursued goal in medical image analysis. We presented a method based on self-supervised learning, which builds an implicit evolution model by taking benefit from longitudinal unlabelled data. This method allows to build representations in an unsupervised way that captures time-specific patterns in the data. The representation can be transferred to solve many longitudinal problems, such as patient-specific early prediction or risk estimation. Unlike reconstruction based methods, the pretext task allows to train on irregular longitudinal data, with irregular time-sampling or limited anatomical registration. The trained encoder can be transferred easily to related longitudinal problems with limited amount of annotated data. In this paper, we applied the method on longitudinal OCTs of patients with intermediate AMD. The learned features were successfully transferred to the problem of predicting incoming disease onset to advanced AMD. There are, however, some limitations in the proposed method. The method is trained on single B-scans instead on the full volume, which highly reduces the memory footprint, but introduces some intermediate steps to generate a patient representation and makes the pretext task harder, as the evolution of each OCT volume might not be uniformly distributed. Although we demonstrated the capability of our approach on retinal OCT scans, the method is not limited to this imaging modality or anatomical region, and may be applied to other longitudinal medical imaging datasets as well. 
\section{Acknowledgements}
This work was funded by the Christian Doppler Research Association, the Austrian Federal Ministry for Digital and Economic Affairs and the National Foundation for Research, Technology and Development. We thank the NVIDIA corporation for a GPU donation.

\bibliographystyle{plain}

\bibliography{samplepaper.bib}

\begin{thebibliography}{10}

\bibitem{bromley1994signature}
Jane Bromley, Isabelle Guyon, Yann LeCun, Eduard S{\"a}ckinger, and Roopak
  Shah.
\newblock Signature verification using a "siamese" time delay neural network.
\newblock In {\em Advances in NIPS}, pages 737--744, 1994.

\bibitem{chen_three-dimensional_2012}
X.~Chen, M.~Niemeijer, L.~Zhang, K.~Lee, M.~D. Abramoff, and M.~Sonka.
\newblock Three-dimensional segmentation of fluid-associated abnormalities in
  retinal {OCT}: Probability constrained graph-search-graph-cut.
\newblock {\em {IEEE} Transactions on Medical Imaging}, 31(8):1521--1531, 2012.

\bibitem{Fujimoto_2016}
James Fujimoto and Eric Swanson.
\newblock The development, commercialization, and impact of optical coherence
  tomography.
\newblock {\em Investigative Ophthalmology \& Visual Science},
  57(9):OCT1--OCT13, 2016.

\bibitem{kingma2014adam}
Diederik~P Kingma and Jimmy Ba.
\newblock Adam: A method for stochastic optimization.
\newblock {\em {arXiv}:1412.6980}, 2014.

\bibitem{DBLP:journals/corr/abs-1708-01246}
Hsin{-}Ying Lee, Jia{-}Bin Huang, Maneesh Singh, and Ming{-}Hsuan Yang.
\newblock Unsupervised representation learning by sorting sequences.
\newblock {\em CoRR}, abs/1708.01246, 2017.

\bibitem{lorenzi_efficient_2015}
Marco Lorenzi, Gabriel Ziegler, Daniel~C. Alexander, and Sebastien Ourselin.
\newblock Efficient {{Gaussian Process}}-{{Based Modelling}} and {{Prediction}}
  of {{Image Time Series}}.
\newblock In {\em Information Processing in Medical Imaging (IPMI)}, volume~24,
  pages 626--637, 2015.

\bibitem{sabuncu_event_2014}
Mert~R. Sabuncu, Jorge~L. {Bernal-Rusiel}, Martin Reuter, Douglas~N. Greve,
  Bruce Fischl, and {Alzheimer's Disease Neuroimaging Initiative}.
\newblock Event time analysis of longitudinal neuroimage data.
\newblock {\em NeuroImage}, 97:9--18, August 2014.

\bibitem{schmidt-erfurth_prediction_2018}
Ursula Schmidt-Erfurth, Sebastian~M. Waldstein, Sophie Klimscha, Amir
  Sadeghipour, Xiaofeng Hu, Bianca~S. Gerendas, Aaron Osborne, and Hrvoje
  Bogunovic.
\newblock Prediction of individual disease conversion in early {AMD} using
  artificial intelligence.
\newblock {\em Investigative Ophthalmology \& Visual Science},
  59(8):3199--3208.

\bibitem{Vogl_2017}
Wolf-Dieter Vogl, Sebastian~M. Waldstein, Bianca~S. Gerendas, Thomas Schlegl,
  Georg Langs, and Ursula Schmidt-Erfurth.
\newblock Analyzing and predicting visual acuity outcomes of anti-{VEGF}
  therapy by a longitudinal mixed effects model of imaging and clinical data.
\newblock {\em Investigative Ophthalmology \& Visual Science}, 58(10):4173,
  2017.

\bibitem{Vogl2017a}
Wolf-Dieter Vogl, Sebastian~M Waldstein, Bianca~S Gerendas, Ursula
  Schmidt-Erfurth, and Georg Langs.
\newblock Predicting macular edema recurrence from spatio-temporal signatures
  in optical coherence tomography images.
\newblock {\em IEEE Transactions on Medical Imaging}, 36(9):1773--1783, sep
  2017.

\end{thebibliography}
\end{document}